\documentclass[prd,showpacs,preprintnumbers,amsmath,amssymb]{revtex4}

\usepackage{graphics}
\usepackage{epsfig}
\usepackage{subfigure}
\usepackage{dcolumn}
\usepackage{bm}
\usepackage{color}

\begin{document}
\title{Hidden-charmonium decays of $Z_c(3900)$ and $Z_c(4025)$ in intermediate meson loops model}

\author{Gang Li\footnote{gli@mail.qfnu.edu.cn}}

\affiliation{ Department of Physics, Qufu Normal University, Qufu
273165, People's Republic of China}


\begin{abstract}
The BESIII collaboration reported an observation of two charged
charmonium-like structure $Z_c^{\pm}(3900)$ and $Z_c^{\pm}(4025)$ in $e^+e^- \to (J/\psi \pi)^{\pm} \pi^{\mp}$ and $e^+e^- \to (D^* {\bar D}^*)^{\pm} \pi^{\mp}$ at ${\sqrt s} =4.26$ GeV recently, which could be an analogue of
$Z_b(10610)$ and $Z_b(10650)$ claimed by the Belle Collaboration.  In this work, we
investigate the hidden-charmonium transitions of $Z_c^{\pm}(3900)$ and $Z_c^{\pm}(4025)$ via
intermediate $D^{(*)} {D}^{(*)}$ meson loops. Reasonable results for
the branching ratios by taking appropriate values of $\alpha$ in
this model can be obtained, which shows that the intermediate
$D^{(*)} {D}^{(*)}$ meson loops process may be a possible mechanism
in these decays. Our results are consistent with the power-counting analysis, and comparable with the calculations in the framework of nonrelativistic effective field theory to some extent. We expect more experimental
measurements on these hidden-charmonium decays and
search for the decays of $Z_c\to
D{\bar D}^* +c.c.$ and $Z_c^\prime \to D^* {\bar D}^*$, which will help us investigate
the $Z_c^{(\prime)}$ decays deeply.
\end{abstract}
\pacs{13.25.GV, 13.75.Lb, 14.40.Pq}

\date{\today}
 \maketitle

\section{Introduction}
\label{sec:introduction}

Recently, a new charged state $Z_c^{\pm}(3900)$ (abbreviated to $Z_c^{\pm}$ in the following) is observed in the $J/\psi \pi^{\pm}$ invariant mass spectrum of $Y(4260)\to J/\psi \pi^+\pi^-$
decay by the BESIII Collaboration~\cite{Ablikim:2013mio}. The
reported mass and width are $M_{Z_c^\pm }= 3899.0\pm 3.6 \pm 4.9$ MeV and $\Gamma_{Z_c^\pm } = 46\pm 11.3 \pm 12.6$ MeV~\cite{Ablikim:2013mio}. Belle Collaboration also observed a
new charged charmonium-like structure in the  $J/\psi \pi^{\pm}$
invariant mass spectrum with $5.2\sigma$ significance, with mass
$M_{Z_c^\pm }= 3894.5\pm 6.6 \pm 4.5 \rm {MeV}$ and width
$\Gamma_{Z_c^\pm } = 63\pm 24 \pm 26 {\rm MeV}$~\cite{Liu:2013dau}.
The observation was
confirmed later on by an analysis based on the CLEO
data at the energy of $4.17$ GeV \cite{Xiao:2013iha}. Very recently, the BESIII
Collaboration reported another new charged structure $Z_c^{\pm}(4025)$ (abbreviated to $Z_c^{\prime \pm}$ in the following) in
$e^+e^- \to (D^*{\bar D}^*)^\pm \pi^{\mp}$ at ${\sqrt s} =4.26$
GeV~\cite{Ablikim:2013emm}. Different from the other hidden-charmonium-like states, such as $X(3872)$, $Y(4260)$ etc., $Z_c$ is
an electric charged state. Such a state, if it exists, need at least
four quarks as minimal constituents, which makes them ideal
candidates for exotic hadrons beyond the conventional $q\bar{q}$
mesons. On the one hand, this chain decay mode reminds us of the
observations of $Z_b(10610)$ and $Z_b(10650)$ in $\Upsilon(5S)\to
Z_b(Z_b')\pi\to
\Upsilon(nS)\pi\pi$~\cite{Collaboration:2011gja,Belle:2011aa}. On
the other hand, the mass of $Z_c$ and $Z_c^\prime$ are in the vicinity of
$D\bar{D}^*+c.c.$ and $D^*{\bar D}^* $, respectively. This similar phenomenon (mechanism) shows that
$Z_c^{\pm}$ and $Z_c^{\prime\pm}$ may be an analogue of $Z_b(10610)$ and $Z_b(10650)$ claimed by the Belle
Collaboration.

Recently, many investigations have been carried out to explain this exotic states~\cite{Wang:2013cya,Guo:2013sya,Chen:2013wca,Faccini:2013lda,Voloshin:2013dpa,Mahajan:2013qja,Cui:2013yva,Wilbring:2013cha}.
The results of Ref.~\cite{Wang:2013cya} show that it is necessary to
include explicit $Z_c(3900)$ poles, i.e. a resonance structure as an
isovector partner of $X(3872)$ in order for a more detailed
description of the data. Applying heavy quark spin symmetry and
heavy flavor symmetry~\cite{Guo:2013sya}, the authors found a
promising isovector $1^{+-}$ $D {\bar D}^*$ virtual state near
threshold that might very well be identified with the newly
discovered $Z_c(3900)$~\cite{Ablikim:2013mio,Liu:2013dau}.

The intermediate meson loop (IML) transitions, or known as final
state interactions, have been one of the important non-perturbative
transition mechanisms in many
processes~\cite{Li:1996yn,Cheng:2004ru,Anisovich:1995zu,Li:2011ssa,Li:2007ky,Wang:2012mf,Li:2007xr,Zhao:2006dv,Zhao:2006cx,Wu:2007jh,Liu:2006dq,Zhao:2005ip,Li:2007au,Zhang:2009kr,Liu:2009vv,Liu:2010um,Wang:2012wj,Guo:2010zk,Guo:2010ak,Zhao:2013jza}.
In the energy region of charmonium masses, with more and more data
from Belle, BaBar, CLEO and BESIII, it is widely recognized that
intermediate hadron loops may be closely related to a lot of
non-perturbative phenomena observed in
experiment~\cite{Wu:2007jh,Liu:2006dq,Cheng:2004ru,Anisovich:1995zu,Zhao:2005ip,Li:2007au,Zhang:2009kr,Liu:2009vv,Liu:2010um,Wang:2012wj,Guo:2010zk,Guo:2010ak,Zhao:2013jza,Brambilla:2010cs,Brambilla:2004wf,Brambilla:2004jw},
e.g. apparent OZI-rule violations, sizeable non-$D\bar D$ decay
branching ratios for $\psi(3770)$, and the helicity selection rule
violations in charmonium decays. Recently, the IML transitions are
also applied to bottomium
decays~\cite{Meng:2007tk,Meng:2008bq,Sun:2011uh,Cleven:2013sq,Li:2012as}.
By applying the on-shell approximation, the bottom meson loops were
suggested to play an important role in the $\Upsilon(5S)$
transitions to the lower $\Upsilon$ states with the emission of two
pions~\cite{Meng:2007tk} or one $\eta$~\cite{Meng:2008bq}. This
mechanism seems to explain many unusual properties that make the
$\Upsilon(5S)$ different from $\Upsilon(4S)$. Similar approach was
also applied to the study of $Z_b$ and $Z_b^\prime$ by Liu {\it et
al.}~\cite{Sun:2011uh}. Within a nonrelativistic effective field
theory (NREFT), the decays of the $Z_b(10610)$ and the $Z_b(10650)$ to
$\Upsilon(nS)\pi$ and $h_b(mP)\pi$ are investigated in
Ref.~\cite{Cleven:2013sq}. The power-counting analysis in
Ref.~\cite{Cleven:2013sq} shows that the triangle transition $Z_b
\to h_b(mP) \pi$ is not suppressed compared to $Z_b \to \Upsilon(nS)
\pi$, although the decay is via a P-wave. In Ref.~\cite{Li:2012as},
we investigated the transitions from the $Z_b(10610)$ and
$Z_b(10650)$ to bottomonium states with emission of a pion via
intermediate $BB^*$ meson loops in the effective Lagrangian approach
(ELA). The results show that the intermediate $B B^*$ meson loops
are crucial for driving the transitions of $Z_b/Z_b^\prime \to
\Upsilon(nS)\pi$ with $n=1,2,3$, and $h_b(mP)\pi$ with $m=1$ and
$2$.

Since the $Z_c^{\pm}$ and $Z_c^{\prime \pm}$ are very close to the $D{\bar D}^*$ and $D^*{\bar D}^*$ thresholds,
the IML should be a possible mechanism in their decays. In this work,
we will investigate the decays of $Z_c^{(\prime)} \to J/\psi \pi$,
$\psi^\prime \pi$ and $h_c\pi$ via intermediate charmed meson loops in
an effective Lagrangian approach (ELA) with quantum numbers
$I^G(J^{PC}) = 1^+ (1^{+-})$ for the $Z_c/Z_c'$. The paper is
organized as follows. In Sec.~\ref{sec:formula}, we will introduce
the formulas for the ELA.  In Sec.~\ref{sec:results}, the numerical
results are presented. A summary will be given in Sec.~\ref{sec:summary}.

\begin{figure}[ht]
\centering
\includegraphics[scale=0.9]{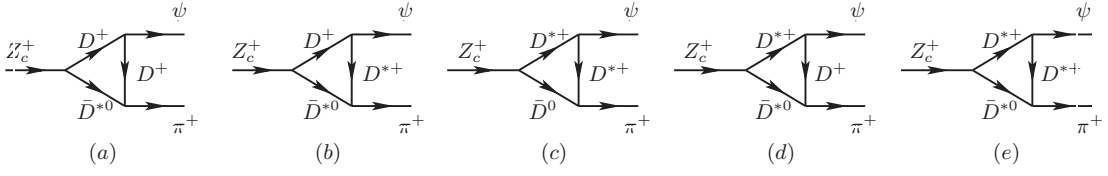}
\caption{Schematic picture for the decay of $Z_c^+ \to \psi \pi^+$
via $D^{(*)} D^{(*)}$ intermediate charmed meson loops. Similar diagrams
for $Z_c^-$ and $Z_c^0$ states decays.}\label{fig:feyn-zc-psi}
\end{figure}

\begin{figure}[ht]
\centering
\includegraphics[scale=0.9]{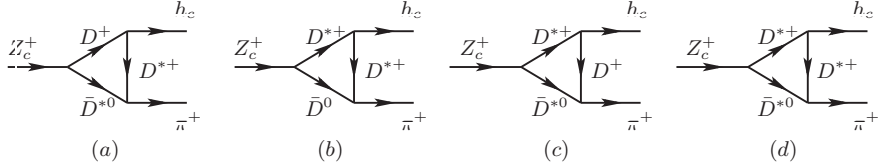}
\caption{Schematic picture for the decay of $Z_c^+ \to h_c \pi^+$
via $D^{(*)} D^{(*)}$ intermediate charmed meson loops. Similar diagrams
for $Z_c^-$ and $Z_c^0$ states decays.}\label{fig:feyn-zc-hc}
\end{figure}
\section{Transition Amplitude}
\label{sec:formula}

The IML transitions can be schematically illustrated in
Figs.~\ref{fig:feyn-zc-psi} and \ref{fig:feyn-zc-hc}. In order to
calculate the leading contributions from the charmed meson loops, we
need the leading order effective Lagrangians for the couplings.
Based on the heavy quark symmetry~\cite{Colangelo:2003sa,Casalbuoni:1996pg}, the relevant
effective Lagrangians used in this work are as follows,
\begin{eqnarray}
\mathcal{L}_1 &=& i g_1 Tr[P_{c\bar{c}}^\mu \bar{H}_{2i}\gamma_\mu
\bar{H}_{1i}] + h.c., \\
\mathcal{L}_2 &=& i g_2 Tr[R_{c\bar{c}} \bar{H}_{2i}\gamma^\mu
{\stackrel{\leftrightarrow}{\partial}}_\mu \bar{H}_{1i}] + h.c.,
\end{eqnarray}
where the spin multiplets for these four $P$-wave and two $S$-wave
charmonium states are expressed as
\begin{eqnarray}
P_{c\bar{c}}^\mu &=& \left( \frac{1+ \rlap{/}{v} }{2} \right)
\left(\chi_{c2}^{\mu\alpha}\gamma_{\alpha} +\frac{1}{\sqrt{2}}
\epsilon^{\mu\nu\alpha\beta}v_{\alpha}\gamma_{\beta}\chi_{c1\nu}
+\frac{1}{\sqrt{3}}(\gamma^\mu -v^\mu)\chi_{c0} +h_c^\mu \gamma_{5}
\right) \left( \frac{1- \rlap{/}{v} }{2} \right), \\
R_{c\bar{c}}&=&  \left( \frac{1+ \rlap{/}{v} }{2} \right) (\psi^\mu
\gamma_\mu-\eta_c \gamma_5) \left( \frac{1- \rlap{/}{v} }{2}
\right).
\end{eqnarray}
The charmed and anti-charmed meson triplet read
\begin{eqnarray}
H_{1i}&=&\left( \frac{1+ \rlap{/}{v} }{2} \right)
[\mathcal{D}_i^{*\mu}
\gamma_\mu -\mathcal{D}_i\gamma_5], \\
H_{2i}&=& [\bar{\mathcal{D}}_i^{*\mu} \gamma_\mu
-\bar{\mathcal{D}}_i\gamma_5]\left( \frac{1- \rlap{/}{v} }{2}
\right),
\end{eqnarray}
where $\mathcal{D}$ and $\mathcal{D}^*$ denote the pseudoscalar and
vector charmed meson fields, respectively, i.e. $\mathcal{D}^{(*)}=(D^{0(*)},D^{+(*)},D_s^{+(*)})$.

Explicitly, the Lagrangians for the S-wave ($J/\psi$ and
$\psi^\prime$) and P-wave ($h_c$) charmonia couplings to $D$ and $D^*$ become
\begin{eqnarray}
\mathcal{L}_{\psi D^{(*)} D^{(*)}} &=&
-ig_{\psi D^* D^*} \big\{
\psi^\mu (\partial_{\mu} D^{* \nu} \bar{D}^*_{\nu}
-D^{* \nu} \partial_{\mu}
\bar{D}^*_{\nu})+ (\partial_{\mu} \psi_{\nu} D^{* \nu} -\psi_{\nu}
\partial_{\mu} D^{* \nu}) \bar{D}^{* \mu} +
D^{* \mu}(\psi^\nu \partial_{\mu} \bar{D}^*_{\nu} -
\partial_{\mu} \psi^\nu \bar{D}^*_{\nu})\big\} \nonumber \\
&& + ig_{\psi DD} \psi_{\mu} (\partial^\mu D \bar{D}- D
\partial^\mu \bar{D})-g_{\psi D^* D} \varepsilon^{\mu \nu
\alpha \beta}
\partial_{\mu} \psi_{\nu} (\partial_{\alpha} D^*_{\beta} \bar{D}
 + D \partial_{\alpha}
\bar{D}^*_{\beta}), \label{eq:h1} \\
\mathcal{L}_{h_c D^{(*)} D^{(*)}}&=& g_{h_c D^*
D} h_c^\mu ( D \bar{D}^*_{\mu}+ D^*_\mu \bar{D})+ ig_{h_c
D^* D^*} \varepsilon^{\mu \nu \alpha \beta}
\partial_{\mu} h_{c \nu} D^*_{\alpha} \bar{D}^*_{\beta} \ .\label{eq:h2}
\end{eqnarray}
The relevant Lagrangians for $Z_c$ and $Z_c^\prime$ couplings to a
pair of charmed mesons can be expressed as
\begin{eqnarray}
\mathcal{L}_{Z_c^{(\prime)} D^{(*)} D^{(*)}}&=& g_{Z_c^{(\prime)}
D^*D} Z_c^{(\prime)\mu} ( D \bar{D}^*_{\mu}+ D^*_\mu \bar{D})+
ig_{Z_c^{(\prime)} D^* D^*} \varepsilon^{\mu \nu \alpha \beta}
\partial_{\mu} Z_{c \nu}^{(\prime)} D^*_{\alpha} \bar{D}^*_{\beta} \ ,\label{eq:h4}
\end{eqnarray}
and the Lagrangian relevant to light pseudoscalar pion meson is
\begin{eqnarray}
{\cal L}_{D^* D^{(*)}\pi} &=& -i g_{D^*D\pi}(D_i \partial_\mu P_{ij} {\bar D}_j^{*\mu} -  D_i^{*\mu} \partial_\mu P_{ij} {\bar D}_j) + \frac {1} {2} g_{D^*D^* \pi} \varepsilon^{\mu\nu\alpha\beta} {D}_{i\mu}^* \partial_\nu P_{ij} {\overleftrightarrow\partial}_\alpha {\bar D}_{j\beta}^*. \label{eq:h3}
\end{eqnarray}
The coupling constants will be determined in the following.

The loop transition amplitudes for the transitions in
Figs.~\ref{fig:feyn-zc-psi} and \ref{fig:feyn-zc-hc} can be
expressed in a general form in the effective Lagrangian approach as
follows:
 \begin{eqnarray}
 M_{fi}=\int \frac {d^4 q_2} {(2\pi)^4} \sum_{D^* \ \mbox{pol.}}
 \frac {V_1V_2V_3} {a_1 a_2 a_3}\prod_i{\cal F}_i(m_i,q_i^2)
 \end{eqnarray}
where $V_i \ (i=1,2,3)$ are the vertex functions; $a_i = q_i^2-m_i^2
\ (i=1,2,3)$ are the denominators of the intermediate meson
propagators. We adopt the form factor, $\prod_i{\cal
F}_i(m_i,q_i^2)$, which is a product of  monopole form factors for
each of the internal mesons, i.e.
\begin{equation}\label{ELA-form-factor}
\prod_i{\cal F}_i(m_i,q_i^2)\equiv \prod_i \frac
{\Lambda_i^2-m_{i}^2} {\Lambda_i^2-q_i^2} ,
\end{equation}
where $\Lambda_i\equiv m_i+\alpha\Lambda_{\rm QCD}$ and the QCD
energy scale $\Lambda_{\rm QCD} = 220$ MeV. This parameter scheme
has been applied extensively in other
works~\cite{Liu:2009vv,Liu:2010um,Guo:2010ak,Li:2012as}. This form
factor is supposed to offset the off-shell effects of the exchanged mesons~\cite{Li:1996yn,Locher:1993cc,Li:1996cj}. In this approach the local couplings for a charmonium to charmed mesons, or a light meson to charmed mesons, are the same as used in NREFT~\cite{Guo:2010ak}, while the form factor parameter will be determined by comparison to experimental information. Thus, it is assumed here that all (at least the dominant part) of the short
range physics related to meson loops can be parameterized in the
form of Eq.~(\ref{ELA-form-factor}).

Based on the above Lagrangians, the explicit amplitudes in
Figs.~\ref{fig:feyn-zc-psi} and \ref{fig:feyn-zc-hc} can be obtained,
which are given in the Appendix~\ref{appendix-A}.

\section{Numerical results}
\label{sec:results}

\begin{figure}[tb]
\begin{center}
\vglue-0mm
\includegraphics[width=0.45\textwidth]{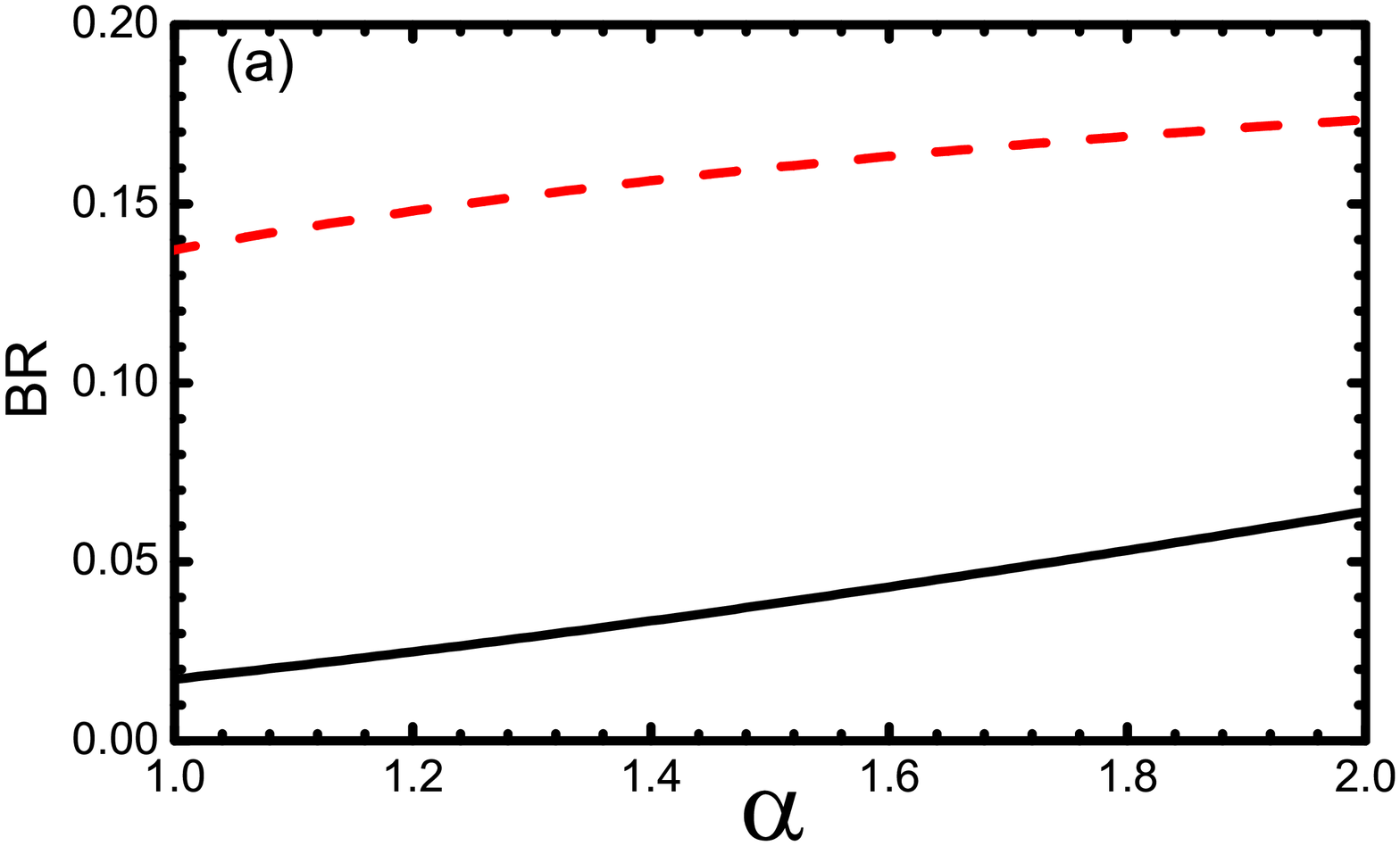}
\includegraphics[width=0.45\textwidth]{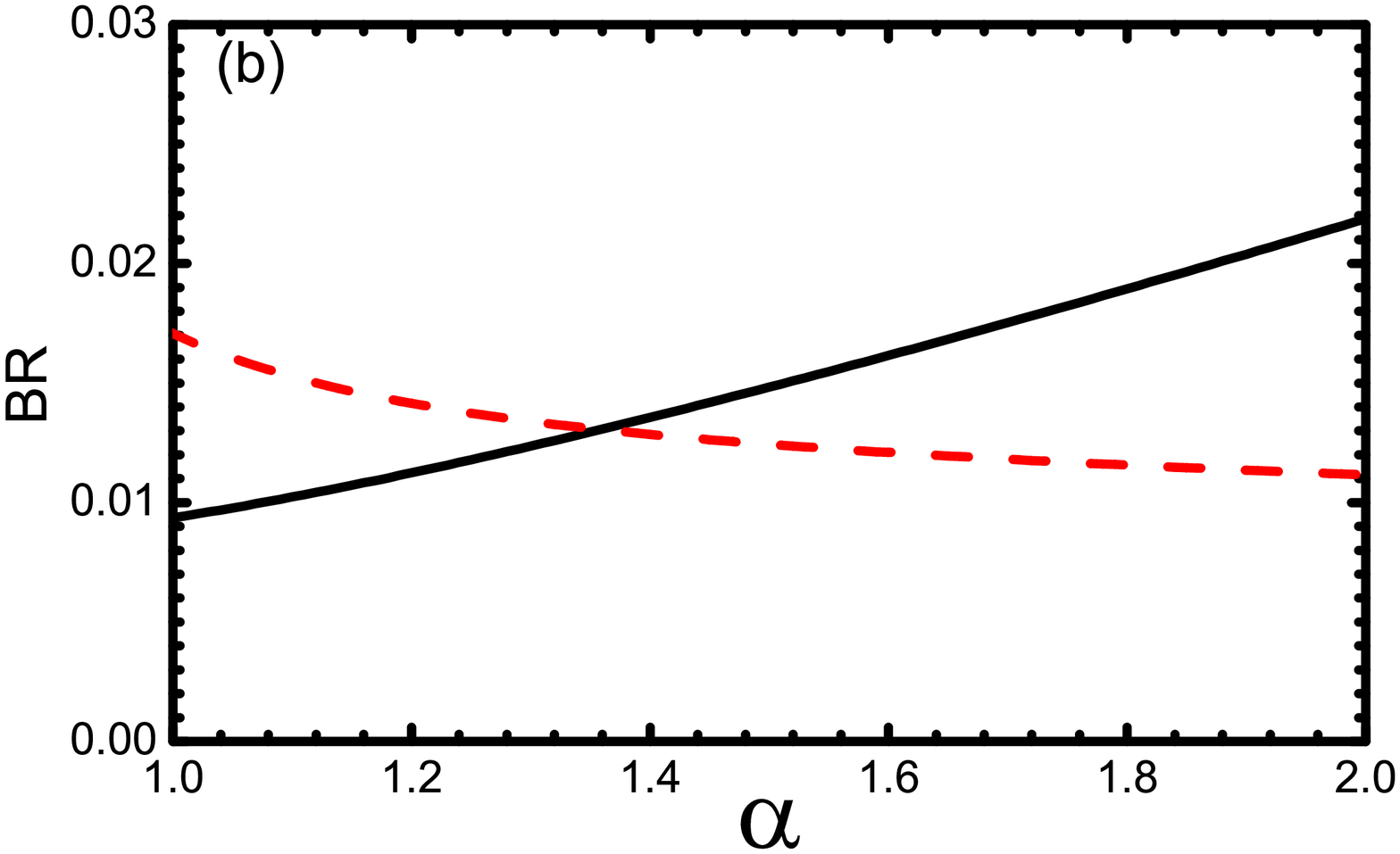}
\caption{ (a) The $\alpha$-dependence of the branching ratios of
$Z_c^+ \to J/\psi \pi^+$ (solid line) and
$\psi^\prime \pi^+$ (dashed line). (b) The $\alpha$-dependence of the branching ratios of
$Z_c^{\prime +} \to J/\psi \pi^+$ (solid line),
$\psi^\prime\pi^+$ (dashed line). }\label{fig:alpha_zcpsipi1}
\end{center}
\end{figure}

\begin{figure}[tb]
\begin{center}
\vglue-0mm
\includegraphics[width=0.45\textwidth]{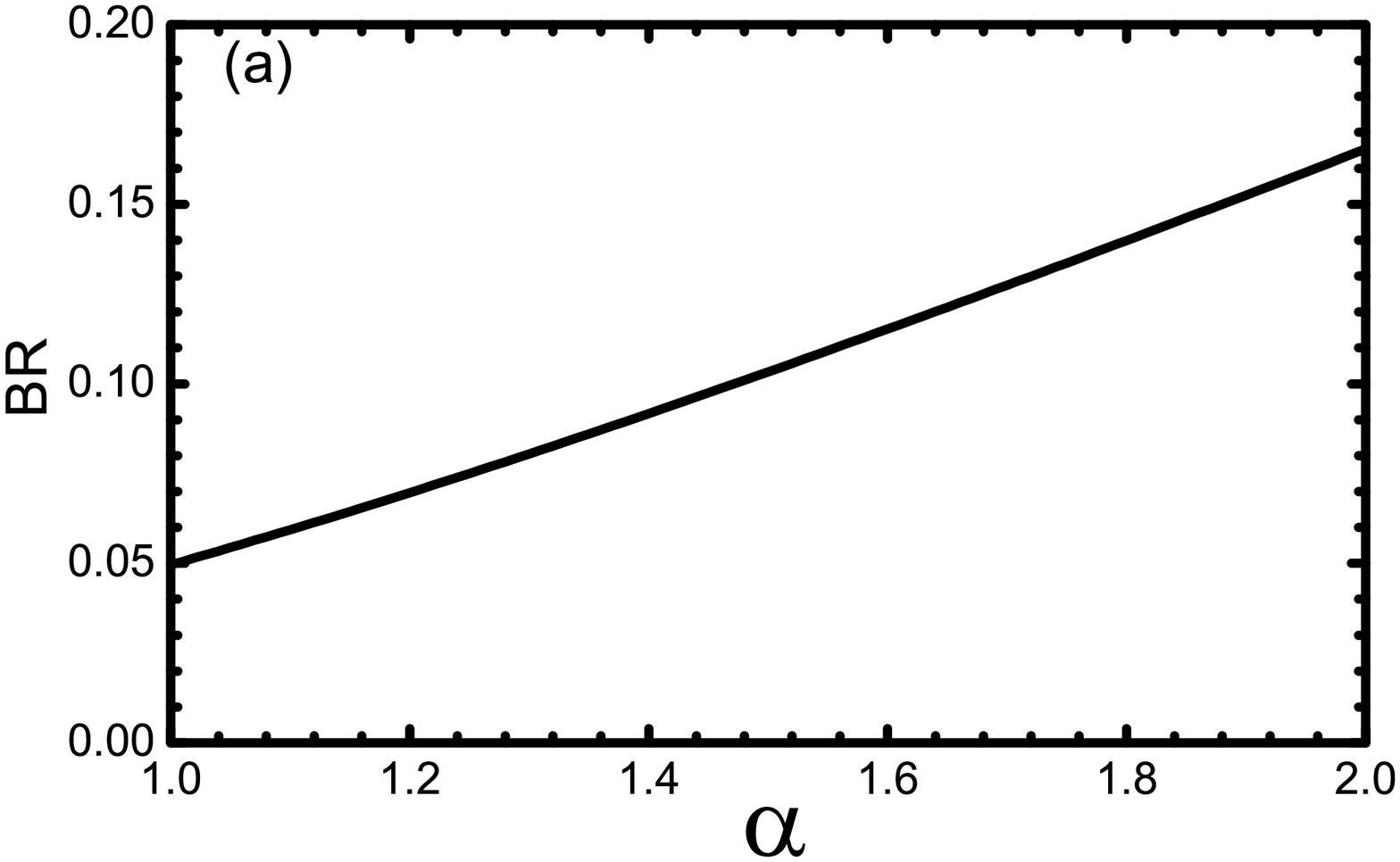}
\includegraphics[width=0.45\textwidth]{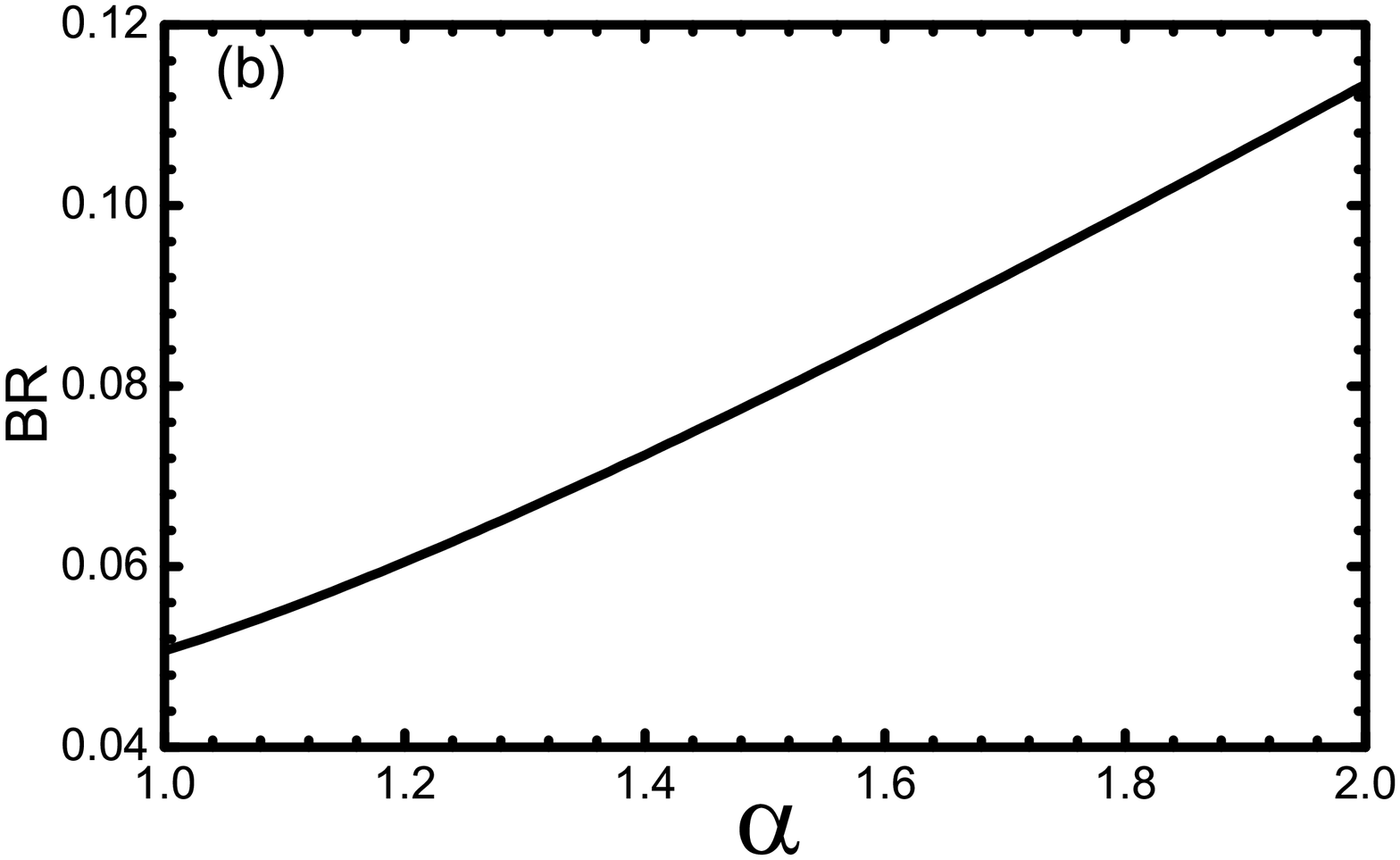}
\caption{ (a) The $\alpha$-dependence of the branching ratios of
$Z_c^+ \to h_c \pi^+$. (b) The $\alpha$-dependence of the branching ratios of
$Z_c^{\prime +} \to h_c \pi^+$.}\label{fig:alpha_zchcpi1}
\end{center}
\end{figure}

\begin{figure}[tb]
\begin{center}
\vglue-0mm
\includegraphics[width=0.45\textwidth]{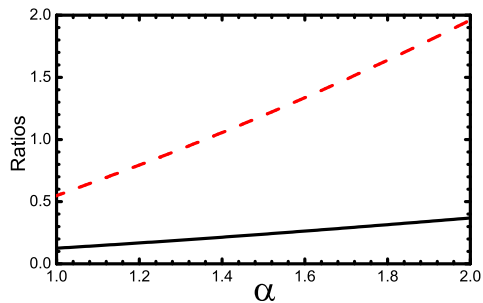}
\caption{ The $\alpha$-dependence of the ratios ${\rm R}_{Z_c}$ (solid line) and ${\rm R}_{Z_c^\prime}$ (dashed line) defined in Eq.~(\ref{eq:ratio-1}). }\label{fig:ratio}
\end{center}
\end{figure}
Before proceeding to the numerical results, we first discuss the
parameters, such as the coupling constants, in the formulation given
in Section.~\ref{sec:formula}. In Eq.~(\ref{eq:h1}), the following
coupling constants are adopted in the numerical calculations,
\begin{eqnarray}
g_{\psi DD} = 2g_2 \sqrt{m_\psi} m_D \ ,
\quad g_{\psi D^* D} = \frac {g_{\psi DD}} {\sqrt{m_D m_{D^*}}} \ ,
\quad g_{\psi D^* D^*} = g_{\psi D^* D}  \sqrt{\frac {m_{D^*}} {m_D}} m_{D^*} \ ,
\end{eqnarray}
In principle, the coupling $g_2$ must be computed by non-perturbative
methods. It shows that vector meson dominance (VMD) would provide an
estimate of these quantities~\cite{Colangelo:2003sa}. The coupling
$g_2$ can be related to the $J/\psi$ leptonic constant $f_\psi$
which is defined by the matrix element $\langle 0| \bar c\gamma_\mu
c|J/\psi (p, \epsilon)\rangle = f_\psi m_\psi \epsilon^\mu$, and
$g_2={\sqrt{m_\psi}}/(2m_Df_\psi),$ where $f_\psi = 405\pm 14$ MeV,
and we have applied the relation $g_{\psi DD} = {m_\psi} /{f_\psi}$.

The ratio of the coupling constants $g_{\psi' DD}$ to $g_{\psi DD}$
is fixed as in Ref.~\cite{Wang:2012wj}:
\begin{eqnarray}
\frac {g_{\psi'DD}} {g_{\psi DD}} = 0.9.
\end{eqnarray}

In addition, the coupling constants in Eq.~(\ref{eq:h2}) are
determined as
\begin{eqnarray}
g_{h_c DD^*} &=& -2g_1 \sqrt{m_{h_c} m_D m_{D^*}} , \ \ g_{h_c
D^* D^*} =2 g_1 \frac{m_{D^*}}{\sqrt{m_{h_c}}},
\end{eqnarray}
with $g_1=-\sqrt{{m_{\chi_{c0}}}/{3}}/{f_{\chi_{c0}}}$, where $m_{\chi_{c0}}$ and $f_{\chi_{c0}}=510 \pm 40$ MeV are the mass and decay constant
of $\chi_{c0}(1P)$, respectively~\cite{Colangelo:2002mj}.

In analogy to what is known about the $Z_b$ states, we assume that
the total widths of the $Z_c^+$ and $Z_c^{\prime +}$ are saturated by the $DD^*$ and $D^*D^*$. The coupling constants are obtained with the following relations
\begin{eqnarray}
g_{Z_c^{(\prime)} DD^*} &=& -2z^{(\prime)} \sqrt{m_{Z_c^{(\prime)}} m_D m_{D^*}} , \quad g_{Z_c^{(\prime)}
D^* D^*} =2 z^{(\prime)} \frac{m_{D^*}}{\sqrt{m_{Z_c^{(\prime)}}}},
\end{eqnarray}
with $z= (0.85^{+0.07}_{-0.26})$ GeV$^{-1/2}$ and $z^\prime=(0.33^{+0.06}_{-0.07})$ GeV$^{-1/2}$.

The coupling constants relevant to the pion interactions in Eq.~(\ref{eq:h4}) are
\begin{eqnarray}
g_{D^* D \pi} = \frac{2
g}{f_\pi} \sqrt{m_D m_{D^*}} \ ,
\quad g_{D^* D^* \pi}=\frac{g_{D^* D \pi}}{\sqrt{m_D m_{D^*}}} \  ,
\end{eqnarray}
where $f_\pi = 132$ MeV is the pion decay constant and $g =
0.59$~\cite{Isola:2003fh}.

In Ref.~\cite{Guo:2009wr}, the NREFT method was introduced to study the meson loop effects in
$\psi^\prime \to J/\psi \pi^0$ transitions. And a power-counting
scheme was proposed to estimate the contribution of the loop
effects, which is helpful to judge how important the coupled-channel
effects are. This power-counting scheme was analyzed in detail in
Ref.~\cite{Guo:2010ak}. Before giving the explicit numerical results, we will follow a
similar power-counting scheme to qualitatively estimate the
contributions of the coupled-channel effects discussed in this work.
Corresponding to the diagrams Figs.~\ref{fig:feyn-zc-psi} and \ref{fig:feyn-zc-hc}, the amplitudes for $Z_c^+/Z_c^{\prime +}\to
J/\psi\pi^+$ ($\psi^\prime \pi^+$)
and $Z_c^+/Z_c^{\prime +}\to h_c\pi^+$ scale as
\begin{eqnarray}
\frac{v^5} {(v^2)^3} q^2  \sim
\frac{q^2}{v} \ \label{power:jpsipi}
\end{eqnarray}
and
\begin{eqnarray}
\frac{v^5} {(v^2)^3} q \sim \frac {q} {v}
\ , \label{power:hcpi}
\end{eqnarray}
respectively. There are two scaling parameters $v$ and $q$ appeared
in the above two formulas. As illustrated in Ref.~\cite{Guo:2012tg},
$v$ is understood as the average velocity of the intermediated
charmed meson. $q$ denotes the momentum of the outgoing pseudoscalar
meson. According to Eqs.~(\ref{power:jpsipi}) and (\ref{power:hcpi}), it can be concluded that the contributions of the coupled-channel effects would be significant here since the amplitudes scale as
$\mathcal{O}(1/v)$. And the branching ratio of $Z_c^+/Z_c^{\prime +} \to
h_c \pi^+$ is expected to be larger than that of $Z_c^+/Z_c^{\prime +} \to
J/\psi \pi^+$, because the corresponding amplitudes scale as
$\mathcal{O}(q)$ and $\mathcal{O}(q^2)$, respectively. However, the
momentum $q$ in $Z_c^+/Z_c^{\prime +} \to J/\psi\pi^+$ is larger than that in
$Z_c^+/Z_c^{\prime +} \to h_c \pi^+$, which may compensate this discrepancy to some
extent.

Since there are no experimental data for the hidden-charmonium
decays of $Z_c^{\pm}$ and $Z_c^{\prime \pm}$, we cannot determine the cutoff parameter $\alpha$
for each channels. However, due the similarity to the hidden-bottom
decays of $Z_b$, it is also possible to find an appropriate range of
$\alpha$ values for each decay channels that can account for the
data via the intermediate charmed meson loops~\cite{Li:2012as}. And
the future experimental measurements can help us test this point. So
in this work, we only present the $\alpha$-dependence of the
hidden-charmonium decays of $Z_c^{\pm}$ and $Z_c^{\prime \pm}$.

In Fig.~\ref{fig:alpha_zcpsipi1}(a), we plot the $\alpha$-dependence
of the branching ratios of $Z_c^+ \to J/\psi \pi^+$ (solid line) and
$\psi'\pi^+$ (dashed line), respectively. The
$\alpha$-dependence is not drastically sensitive at the commonly accepted $\alpha$
range. As shown in this
figure, at the same $\alpha$, the intermediate D-meson loop effects
turn out to be more important in $Z_c^+ \to \psi^\prime \pi^+$ than in
$Z_c^+ \to J/\psi \pi^+$. This is understandable since
the mass of $\psi^\prime$ is closer to the thresholds of $DD^*$ or
$D^*D^*$ than $J/\psi$~\cite{Beringer:1900zz}. Thus, it gives rise
to important threshold effects in $Z_c^+ \to \psi^\prime \pi^+$.

One also notices that the $\alpha$-dependence of the branching ratios
for $Z_c^+ \to \psi^\prime \pi^+$ are less sensitive than that for $J/\psi
\pi^+$. This indicates that the enhanced branching ratios are not from
the off-shell part of the loop integrals and the enhanced (but rather stable in terms of $\alpha$)
branching ratios for $Z_c^+ \to \psi^\prime \pi^+$ suggest that more
stringent dynamic constraints are presumably needed to describe the
near-threshold phenomena where the local quark-hadron duality has
been apparently violated. What makes this process different from
e.g. $\psi'\to h_c\pi^0$ in Ref.~\cite{Guo:2010zk} is that there is
no cancelations between the charged and neutral meson loops. As a
consequence, the subleading terms in
Refs.~\cite{Guo:2010zk,Guo:2010ak} become actually leading
contributions. In Fig.~\ref{fig:alpha_zcpsipi1}(b), we plot the $\alpha$-dependence of
the branching ratios of $Z_c^{\prime +} \to J/\psi \pi^+$ (solid line)
and $\psi^\prime \pi^+$ (dashed line), respectively.
At the commonly accepted $\alpha$ range, the $\alpha$-dependence of the branching ratios is not dramatically sensitive.

The $\alpha$-dependence of the branching ratios of $Z_c^{+} \to h_c\pi^+$ and $Z_c^{\prime +} \to h_c\pi^+$
are shown in Figs.~\ref{fig:alpha_zchcpi1} (a) and (b). From this figure, we
can see that the intermediate meson loop contributions are more important
in $Z_c^{\prime +} \to h_c\pi^+$ than that in $Z_c^{+} \to h_c\pi^+$.

It would be interesting to further clarify the uncertainties arising
from the introduction of form factors by studying the $\alpha$
dependence of the ratios between different partial decay widths. For
the decays of $Z_c^+/Z_c^{\prime +} \to J/\psi \pi^+$, we define the
following ratios to the partial decay widths of $Z_c^+/Z_c^{\prime
+}\to \psi'\pi^+$:
\begin{eqnarray}
{\rm R}_{Z_c}=\frac {\Gamma(Z_c^+ \to J/\psi \pi^+)} {\Gamma(Z_c^+ \to \psi' \pi^+)} , \quad
{\rm R}_{Z_c^\prime}=\frac {\Gamma(Z_c^{\prime +} \to J/\psi \pi^+)} {\Gamma(Z_c^{\prime +}\to \psi^\prime \pi^+)} ,\label{eq:ratio-1}
\end{eqnarray}
which are plotted in Fig.~\ref{fig:ratio}. The ratios are relatively insensitive to the cutoff parameter, which is because the involved loops are the same.
Since the first coupling vertices are the same for those decay channels when taking the
ratio, also the mass of $\psi^\prime$ is closer to $D^{(*)}D^{(*)}$ thresholds than $J/\psi$, so the ratio
only reflects the open threshold effects via the intermediate charmed
meson loops. The future experimental measurements of $Z_c^+/Z_c^{\prime +}
\to J/\psi \pi^+$ and $\psi^\prime \pi^+$ can help us investigate this issue deeply.

In order to understand this, the following analysis is carried out.
First, one notices that we have adopted the couplings for the $h_c$
and $\psi$ to $D\bar{D}^*$ or $D^*\bar{D}^*$ in the heavy quark
approximation. Since the physical masses for $D$ and $D^*$ are
adopted in the loop integrals, the form factor will introduce
unphysical pole contributions of which the interferences with the
nearby physical poles would lead to model-dependent uncertainties.
By assuming $M_{D^*} = M_D= 1869$ MeV and $M_{D^*} = M_D=2010$ MeV,
namely, by making the spin symmetry exactly, we
calculate the branching ratios of $Z_c^+$ and $Z_c^{\prime +}\to
J/\psi\pi^+$, $\psi^\prime \pi^+$ and $h_c\pi^+$. We expect that the
exact spin symmetry will
significantly lower the branching ratios since there will be only one
physical pole in the loop and the unphysical one can be easily
isolated away from the physical one. This is a rather direct
demonstration of the sensitivity of the meson loop behavior when
close to open threshold and when the dispersive part becomes
dominant.

As a cross-check, we also calculate the branching ratios of the decays in the framework of NREFT and the relevant transition amplitudes are given in Appendix~\ref{appendix-B}. The numerical results in NREFT are listed in Table.~\ref{tab:br}, the uncertainties are due to the experimental measured uncertainties of the total widths of $Z_c^{\pm}$ and $Z_c^{\prime \pm}$.
As shown in this table, the results calculated in NREFT are in good agreement with the results in ELA at the commonly accepted range except for the case of $Z_c^+ \to J/\psi \pi^+$, which indicates the availability of our model to some extent and shows that the higher order effects are important in $Z_c^+ \to J/\psi \pi^+$. In fact, following the NREFT power counting for higher loops in Ref.~\cite{Cleven:2013sq}, one sees that this is exactly the channel where higher order loops can be important.
However, since there are still several uncertainties
coming from the undetermined coupling constants, and the cutoff energy dependence of the amplitude is not quite
stable, the numerical results would be lacking in high accuracy. Especially, since the kinematics and off-shell effects arising from the exchanged particles are different, the cutoff parameter can also be different in different decay channels. We expect more experimental
measurements on these hidden-charmonium decays in the near future.

\begin{table}
\begin{center}
\caption{The calculated branching ratios in NREFT approach. The uncertainties are from the experimental measurements uncertainties of the total widths of $Z_c^+$ and $Z_c^{\prime +}$.}\label{tab:br}
\begin{tabular}{cccc}
\hline
              & $Z_c^+$ & $Z_c^{\prime +}$  \\
              & Branching ratios & Branching ratios \\
\hline
$J/\psi \pi^+$  & $(49.89^{+8.82}_{-25.88})\%$ & $(3.21^{+1.25}_{-1.23})\%$ & \\
$\psi^\prime \pi^+$ & $(13.80^{+2.47}_{-7.16})\%$ & $(1.99^{+0.77}_{-0.76})\%$ &\\
$h_c \pi^+$ & $(5.56^{+0.99}_{-2.88})\%$ & $(6.37^{+2.46}_{-2.44})\%$ & \\
\hline
\end{tabular}
\end{center}
\end{table}
\section{Summary}\label{sec:summary}

In this work, we investigate hidden-charm decays of the newly
discovered resonances $Z_c^{\pm}$ and $Z_c^{\prime \pm}$ via intermediate charmed meson loops. In
this calculation, the quantum numbers of the neutral partners of
these two resonances are fixed to be $I^G (J^{PC}) = 1^+(1^{+-})$,
which has the same favored quantum number of $Z_b(10610)$ and $Z_b(10650)$. For $Z_c^{\pm}$ and $Z_c^{\prime \pm}$
decays, our results show that the $\alpha$ dependence of the branching ratios are not dramatically sensitive. Our results show that the meson
loop contributions are much more important when the final state mass threshold are close to
the intermediate meson thresholds. Namely, the effects from the
unphysical pole introduced by the form factors would interfere with
the nearby physical poles from the internal propagators and lead to
model-dependent uncertainties. It is also a consequence of the
violation of spin symmetry and such a phenomenon has been
discussed in Ref.~\cite{Guo:2010ak}. Our results are in good agreement with the results in the framework of NREFT except for the case of $Z_c^+\to J/\psi \pi^+$,  which indicates the availability of our model to some extent.
However, since there are still several uncertainties, for example, the kinematics and off-shell effects
arising from the exchanged particles are
different, the cutoff parameter can also be different in different decay channels,
so we expect more experimental
measurements on these hidden-charmonium decays and
search for the decays of $Z_c\to
D{\bar D}^* +c.c.$ and $Z_c^\prime \to D^* {\bar D}^*$, which will help us investigate
the $Z_c^{(\prime)}$ decays deeply.
\section*{Acknowledgements}

Author thanks F.-K. Guo, X.-H. Liu, Q. Wang and Q. Zhao for useful
discussions. This work is supported, in part, by the National
Natural Science Foundation of China (Grant No. 11275113).

\begin{appendix}

\section{The Transition Amplitude in ELA}
\label{appendix-A}

In the following, we present the transition amplitudes for the
intermediate meson loops listed in Figs.~\ref{fig:feyn-zc-psi} and
\ref{fig:feyn-zc-hc} in the framework of the ELA. Notice that the
expressions are similar for the charged and neutral charmed mesons
except that different charmed meson masses are applied. We thus only
present the amplitudes for those charged charmed meson loops.

(i). $Z_c^{(\prime) +} \to J/\psi \pi^+$ and $\psi^\prime \pi^+$
\begin{eqnarray}
M_{DD^* [D]} &=& (i)^3\int \frac {d^4q_2} {(2\pi)^4}[g_{Z_c^{(\prime)} D^*D}
\varepsilon_{i\mu}] [g_{\psi DD} \varepsilon_f^{*\rho} (q_1-q_2)_\rho] [g_{D^*D\pi}
p_{\pi\theta}] \nonumber \\
&& \frac {i} {q_1^2-m_1^2}  \frac {i} {q_2^2-m_2^2}  \frac
{i(-g^{\mu\theta} +q_3^\mu q_3^\theta/m_3^2)} {q_3^2-m_3^2} \prod_i{\cal F}_i(m_i,q_i^2) \nonumber \\
M_{DD^* [D^*]}&=& (i)^3\int \frac {d^4q_2} {(2\pi)^4}[g_{Z_c^{(\prime)} D^*D}
\varepsilon_{i\mu} ] [g_{\psi D^*D} \varepsilon_{\rho\sigma \xi\tau}p_f^\rho
\varepsilon_f^{*\sigma} q_2^\xi ] [-g_{D^*D^*\pi}
\varepsilon_{\theta\phi\kappa\lambda}  p_{\pi}^\kappa q_2^\lambda]
\nonumber \\
&& \times \frac {i} {q_1^2-m_1^2}  \frac {i(-g^{\tau\theta}
+q_2^\tau q_2^\theta/m_2^2)} {q_2^2-m_2^2}  \frac {i(-g^{\mu\phi}
+q_3^\mu q_3^\phi/m_3^2)} {q_3^2-m_3^2} \prod_i{\cal F}_i(m_i,q_i^2) \nonumber \\
M_{D^*D [D^*]} &=& (i)^3\int \frac {d^4q_2} {(2\pi)^4}[g_{Z_c^{(\prime)} D^*D}
\varepsilon_{i\mu}] [g_{\psi D^*D^*} (g_{\rho\sigma} g_{\xi\tau} - g_{\rho\tau}
g_{\sigma\xi} + g_{\rho\xi} g_{\sigma\tau}) \varepsilon_f^{*\rho}
(q_1+q_2)^\tau] [-g_{B^*B\pi} p_{\pi\theta}] \nonumber \\
&& \times \frac {i(-g^{\mu\xi} +q_1^\mu q_1^\xi/m_1^2)}
{q_1^2-m_1^2}  \frac {i(-g^{\sigma\theta} +q_2^\sigma
q_2^\theta/m_2^2)} {q_2^2-m_2^2}  \frac {i} {q_3^2-m_3^2} \prod_i{\cal F}_i(m_i,q_i^2)  \nonumber \\
M_{D^*D^* [D]} &=& (i)^3\int \frac {d^4q_2} {(2\pi)^4}[g_{Z_c^{(\prime)} D^*D^*}
\varepsilon_{\mu\nu\alpha\beta} q_i^\mu\varepsilon_{i}^\nu] [g_{\psi D^*D}
\varepsilon_{\rho\sigma\xi\tau} p_f^\rho \varepsilon_f^{*\sigma}
q_1^\xi ] [g_{D^*D\pi} p_{\pi\theta}] \nonumber \\
&& \times \frac {i(-g^{\alpha\tau} +q_1^\alpha q_1^\tau/m_1^2)}
{q_1^2-m_1^2}  \frac {i} {q_2^2-m_2^2}  \frac {i(-g^{\beta\theta}
+q_3^\beta q_3^\theta/m_3^2)} {q_3^2-m_3^2} \prod_i{\cal F}_i(m_i,q_i^2) \nonumber \\
M_{D^*D^* [D^*]} &=& (i)^3\int \frac {d^4q_2} {(2\pi)^4}[g_{Z_c^{(\prime)} D^*D^*}
\varepsilon_{\mu\nu\alpha\beta} p_i^\mu\varepsilon_{i}^\nu ] [g_{\psi D^*D^*}
(g_{\rho\sigma} g_{\xi\tau} - g_{\rho\tau} g_{\sigma\xi} +
g_{\rho\xi} g_{\sigma\tau}) \varepsilon_f^{*\rho}  (q_1+q_2)^\tau]
[-g_{D^*D^*\pi}  \varepsilon_{\theta\phi\kappa\lambda} p_{\pi}^\kappa
q_2^\lambda] \nonumber \\ && \times \frac {i(-g^{\alpha\xi}
+q_1^\alpha q_1^\xi/m_1^2)} {q_1^2-m_1^2}  \frac
{i(-g^{\sigma\theta} +q_2^\sigma q_2^\theta/m_2^2)} {q_2^2-m_2^2}
\frac {i(-g^{\beta\phi} +q_3^\beta q_3^\phi/m_3^2)} {q_3^2-m_3^2}
\prod_i{\cal F}_i(m_i,q_i^2) \ .
\end{eqnarray}

(ii). $Z_c^{(\prime) +}\to h_c \pi^+$
\begin{eqnarray}
M_{DD^* [D^*]} &=& (i)^3\int \frac {d^4q_2} {(2\pi)^4}[g_{Z_c^{(\prime)} D^*D} \varepsilon_{i\mu} ] [g_{h_c D^*D} \varepsilon_{f\rho}^{*} ] [-g_{D^*D^*\pi}  \varepsilon_{\theta\phi\kappa\lambda} p_{\pi}^\kappa q_2^\lambda] \nonumber \\
&& \times \frac {i} {q_1^2-m_1^2}  \frac {i(-g^{\rho\theta} +q_2^\rho q_2^\theta/m_2^2)} {q_2^2-m_2^2}  \frac {i(-g^{\mu\phi} +q_3^\mu q_3^\phi/m_3^2)} {q_3^2-m_3^2} \prod_i{\cal F}_i(m_i,q_i^2) \nonumber \\
M_{D^*D [D^*]} &=& (i)^3\int \frac {d^4q_2} {(2\pi)^4}[g_{Z_c^{(\prime)} D^*D} \varepsilon_{i\mu}] [g_{h_c D^*D^*} \varepsilon_{\rho\sigma\xi\tau} p_f^\rho \varepsilon_f^{*\sigma} ] [-g_{D^*D\pi}  p_{\pi\theta} ] \nonumber \\
&& \times \frac {i(-g^{\mu\xi} +q_1^\mu q_1^\xi/m_1^2)} {q_1^2-m_1^2}  \frac {i(-g^{\tau\theta} +q_2^\tau q_2^\theta/m_2^2)} {q_2^2-m_2^2}  \frac {i} {q_3^2-m_3^2} \prod_i{\cal F}_i(m_i,q_i^2) \nonumber \\
M_{D^*D^* [D]} &=& (i)^3\int \frac {d^4q_2} {(2\pi)^4}[g_{Z_c^{(\prime)} D^*D^*} \varepsilon_{\mu\nu\alpha\beta} p_i^\mu\varepsilon_{0}^\nu ] [g_{h_c D^*D} \varepsilon_{f\rho}^{*}] [g_{D^*D\pi}  p_{\pi\theta}] \nonumber \\
&& \times \frac {i(-g^{\alpha\rho} +q_1^\alpha q_1^\rho/m_1^2)} {q_1^2-m_1^2}  \frac {i} {q_2^2-m_2^2}  \frac {i(-g^{\beta\theta} +q_3^\beta q_3^\theta/m_3^2)} {q_3^2-m_3^2} \prod_i{\cal F}_i(m_i,q_i^2)  \nonumber \\
M_{D^*D^* [D^*]} &=& (i)^3\int \frac {d^4q_2} {(2\pi)^4}[g_{Z_c^{(\prime)} D^*D^*} \varepsilon_{\mu\nu\alpha\beta} p_i^\mu\varepsilon_{i}^\nu ] [g_{h_c D^*D^*} \varepsilon_{\rho\sigma\xi\tau} p_f^\rho \varepsilon_f^{*\sigma} ] [-g_{D^*D^*\pi}   \varepsilon_{\theta\phi\kappa\lambda} p_{\pi}^{\kappa} q_2^\lambda] \nonumber \\
&& \times \frac {i(-g^{\alpha\xi} +q_1^\alpha q_1^\xi/m_1^2)}
{q_1^2-m_1^2}  \frac {i(-g^{\tau\theta} +q_2^\tau q_2^\theta/m_2^2)}
{q_2^2-m_2^2}  \frac {i(-g^{\beta\phi} +q_3^\beta q_3^\phi/m_3^2)}
{q_3^2-m_3^2} \prod_i{\cal F}_i(m_i,q_i^2),
\end{eqnarray}
where $p_i$, $p_f$, $p_\pi$ are the four-vector momenta of the
initial $Z_c^{(\prime)}$, final state charmonium and pion, respectively, and
$q_1$, $q_2$, and $q_3$ are the four-vector momenta of the
intermediate charmed mesons as defined in
Figs.~\ref{fig:feyn-zc-psi} and \ref{fig:feyn-zc-hc}.

\section{Amplitudes in NREFT Approach}
\label{appendix-B}

The basic three-point loop function worked out using dimensional regularization in $d=4$ is
\begin{eqnarray}
I(q,m_1,m_2,m_3) &=& \frac{-i}{8} \int \frac{d^d l}{(2\pi)^d} \frac {1} {[l^0 - \frac {\vec {l}\,^2} {m_1}+i\epsilon]} \frac{1} {[l^0-b_{12}+ \frac {\vec {l}\,^2} {m_2}-i\epsilon ]} \frac {1} {[l^0+b_{12}-b_{23}- \frac {(\vec {l}-\vec {q})^2} {m_2}+i\epsilon ]} \nonumber \\
&=& \frac {\mu_{12} \mu_{23}} {16\pi} \frac {1} {{\sqrt 2}} [ \tan^{-1} (\frac {c^\prime -c} {2\sqrt {a(c-i\epsilon)}})  + \tan^{-1} (\frac {2a+c-c^\prime} {2\sqrt {a(c^\prime -a -i\epsilon)}})],
\end{eqnarray}
where $m_i(i=1,2,3)$ are the masses of the particles in the loop; $\mu_{ij}=m_im_j/(m_i+m_j)$ are the reduced masses; $b_{12} =m_1+m_2-M$ and $b_{23} = m_2+m_3+q^0-M$ with M being the mass of the initial particle; and
\begin{eqnarray}
a= (\frac {\mu_{23}} {m_3})^2 {\vec q}\,^2, \quad c=2\mu_{12}b_{12}, \quad c^\prime = 2\mu_{23} b_{23}+\frac {\mu_{23}} {m_3} {\vec q}\,^2.
\end{eqnarray}

The vector loop integrals are defined as
\begin{eqnarray}
q^iI^{(1)}(q,m_1,m_2,m_3) = \frac{-i}{8} \int \frac{d^d l}{(2\pi)^d} \frac {l^i} {[l^0 - \frac {\vec {l}\,^2} {m_1}+i\epsilon] [l^0-b_{12}+ \frac {\vec {l}\,^2} {m_2}-i\epsilon ][l^0+b_{12}-b_{23}- \frac {(\vec {l}-\vec {q})^2} {m_2}+i\epsilon ]}
\end{eqnarray}
and we get
\begin{eqnarray}
I^{(1)}(q,m_1,m_2,m_3) =\frac {\mu_{23}} {am_3} [B(c^\prime-a) -B(c) + \frac {1} {2} (c^\prime -c) I(q)],
\end{eqnarray}
where the function $B(c)$ is
\begin{eqnarray}
B(c)= -\frac {\mu_{12} {\mu_{23}} {\sqrt {c-i\epsilon} }}{16\pi}.
\end{eqnarray}

In terms of the loop functions given above, the transition amplitudes for the
intermediate meson loops listed in Figs.~\ref{fig:feyn-zc-psi} and
\ref{fig:feyn-zc-hc} in the framework of NREFT,

(i) $Z_c^{(\prime) +} \to J/\psi \pi^+$ and $\psi^\prime \pi^+$
\begin{eqnarray}
{\cal M} (Z_c^{(\prime) +}  &\to&  \psi \pi^+) \nonumber \\
&=& -\frac {2{\sqrt 2} g g_1z^{(\prime)}} {f_\pi} {\sqrt {M_{Z_c^{(\prime)}} {M_\psi}}} \{ {\vec q} \cdot {\vec \varepsilon}(Z_c)  {\vec q} \cdot {\vec \varepsilon}(\psi)[2I^{(1)}(q,M_{D^*},M_D,M_D)-I(q,M_{D^*},M_D,M_D) \nonumber \\ &&- 2I^{(1)}(q,M_{D^*},M_D,M_{D^*})
+ I(q,M_{D^*},M_D,M_{D^*})+2I^{(1)}(q,M_{D^*},M_{D^*},M_D)\nonumber \\
&& -I(q,M_{D^*},M_{D^*},M_D)  -2I^{(1)}(q,M_{D^*},M_{D^*},M_{D^*})
+I(q,M_{D^*},M_{D^*},M_{D^*})] \nonumber \\
&& +{\vec q}\ ^2{\vec \varepsilon}(Z_c)\cdot {\vec \varepsilon}(\psi) [2I^{(1)}(q,M_{D^*},M_D,M_{D^*})-I(q,M_{D^*},M_D,M_{D^*}) \nonumber \\ &&+ 2I^{(1)}(q,M_D,M_{D^*},M_{D^*})-I(q,M_D,M_{D^*},M_{D^*})-2I^{(1)}(q,M_{D^*},M_{D^*},M_D)\nonumber \\ && +I(q,M_{D^*},M_{D^*},M_D) - 2I^{(1)}(q,M_{D^*},M_{D^*},M_{D^*})+I(q,M_{D^*},M_{D^*},M_{D^*})]\}
\end{eqnarray}

(ii) $Z_c^{(\prime) +} \to h_c\pi^+$
\begin{eqnarray}
{\cal M}(Z_c^{(\prime) +}\to h_c\pi^+) &=& \frac {2{\sqrt 2}g g_1 z^{(\prime) }} {f_\pi} {\sqrt {M_{Z_c^{(\prime)}} {M_{h_c}}}}\epsilon^{ijk} q^i \varepsilon^j(Z_c) \varepsilon^k(h_c) [I(q, M_D, M_{D^*},M_{D^*})+ I(q, M_{D^*}, M_D,M_{D^*}) \nonumber \\
&& -I(q, M_{D^*}, M_{D^*},M_D)+I(q, M_{D^*}, M_{D^*},M_{D^*})]
\end{eqnarray}
\end{appendix}

\end{document}